*Perspective*

# The first major transition: Origin of life from a multilevel selection perspective

Eugene V. Koonin[1,*]

[1] Computational Biology Branch, Division of Intramural Research, National Library of Medicine, National Institutes of Health of the USA, Bethesda, MD 20894, USA.

For correspondence: koonin@ncbi.nlm.nih.gov

**Abstract**

The origin of life is the first, arguably, the most important and also the most enigmatic major transition in evolution (MTE). As all MTE, the transition at the origin of life can be constructively addressed only within the conceptual framework of multilevel selection. The two levels of selection relevant for the origin of life are replicators, and reproducers, that is, compartments (protocells) capable of reproduction via a primitive form of division. The symbiotic model for the origin of life presented here posits that cellular life was preceded by protocells that harbored proto-metabolic networks but not genetic elements (replicators), Replicators could have started as selfish elements inside protocells selected only for replication efficiency, but subsequently, would evolve into mutualists such that the reproducer-replicator union became a new, collective unit of selection. Mathematical modeling of replicator-reproducer coevolution identifies the conditions under which replicator-carrying protocells can outcompete 'empty' protocells. For the replicator-carrying protocells to and get fixed in evolution, replication has to be coupled with protocell division, selection for the highest replication rate being partially suppressed. At the earliest stages of evolution, random, stochastic protocell division was advantageous compared to symmetrical division. The multilevel selection perspective illuminates the likely order of key events on the evolutionary route from ensembles of chemicals to cells endowed with genomes, symmetrical cell division and anti-parasite defense systems.

**Major transitions in evolution, evolutionary transitions in individuality and multilevel selection**

Evolutionary transition in individuality (ETI) is an important phenomenon in the evolution of life whereby independent reproducing, evolving entities, or Darwinian individuals, combine to form a collective unit, so that a new, more complex type of Darwinian individuals emerges (1-6). In particular, ETI is an essential aspect of major transitions in evolution (MTE), key events in the history of life. The MTEs, as defined by Maynard Smith and Szathmary (3) include: i) origin of cells, ii) origin of eukaryotic cells, iii) origin(s) of multicellular organisms, iv) origin(s) of eusocial animals, v) origin(s) of societies. In addition to the MTEs, ETI accompanied many other evolutionary events, in particular, all cases of symbiosis which abound in the history of life (7-10). In each case of ETI, and MTE in particular, the pressure of selection is mostly transferred to the collectives whereas the selection at the lower level of partner entities is largely suppressed. The case of multicellularity is probably the most intuitive one: in multicellular organisms, proliferation of individual cells is tightly controlled, and impairment of this control can result in severe consequences for the higher-level individuals, such as cancer. To frame the MTE concept more specifically, it is useful to distinguish between fraternal and egalitarian transitions (6, 11). Fraternal transitions are those that involve collectivization of related entities such as cells forming a multicellular organism. Notably, such transitions can be relatively 'easy' as demonstrated by the fact that both multicellularity and eusociality emerged on multiple, independent occasions. Egalitarian transitions involve symbiosis between independent entities of different kinds as in the case of the mutualistic symbiosis between an Asgard archaeon and an alpha-proteobacterium at the origin of eukaryotic cells (12, 13). In egalitarian transitions, the 'interests' of both partners are also aligned and partially suppressed such that the collective unit becomes the principal agency of selection. Egalitarian transitions appear to be even more common in evolution than fraternal ones because all mutualistic symbioses qualify, and these are ubiquitous in the biosphere.

The concept of MTE is tightly connected with the theory of multilevel selection (MLS). According to MLS, selection operates simultaneously at multiple levels, from genes to cells to organisms to populations to, possibly, communities of populations (14-17). This is a traditionally controversial subject tied to the notorious group selection debates in which the opponents of MLS often claim absence of empirical evidence of group selection and, by extrapolation, futility of the MLS concept (18). An objective overview of empirical studies shows, however, that evidence of MLS abounds (19, 20). As a clear-cut, striking case of population-level selection, it is worth pointing out, now extensive, evidence of programmed cell death in prokaryotes that is induced via multiple, dedicated mechanisms as the last line of defense against virus infection or as stress response (21-27). Programmed cell death in unicellular life forms hardly can be explained by any mechanism other than population-level selection, regardless of whether the exact scenario is presented as kin selection or group selection. This and other examples showcase selection at super-organismal levels. The cases for sub-organismal selection are perhaps even more compelling. Indeed, even setting aside the ‘selfish gene’ concept (28), all organisms host multiple mobile genetic elements (MGE), such as transposons, integrating conjugating elements, latent viruses, plasmids, and more, and all have evolved multiple means to control MGE proliferation (29-32). The animal piRNA system is a well-studied example of such a control mechanism (33). Perhaps, MLS is best depicted as two triangles conjoined in a rhombus shape, with the organisms as the fundamental level of selection, with supra-organismal levels ascending, and sub-organismal levels descending (Figure 1).

The ETI and MLS are inseparable because, upon the emergence of a collective unit subject to selection, the selection at the lower level is partially suppressed but not eliminated. All organisms are the scene of intraorganismal and intragenomic conflicts, or put another way, frustrated interactions between the pre-ETI and post-ETI Darwinian individuals (5, 34-37). Such frustrations appear to be a key driver of the evolution of biological complexity (38).

In this Perspective article, I examine the first and most fundamental MTE that is associated with the origin of cells which in this context I consider to be synonymous with the origin of life. This transition is substantially different from all subsequent MTE and other transitions because, in all these later transitions, whether fraternal or egalitarian, the partners are established biological entities and selection is in full swing. Thus, all the fascinating details notwithstanding, compared to the origin of life, those transitions are 'business as usual'. The challenge faced by any attempts to understand is qualitatively harder because, in this case, the transition was from chemical entities, where the possibility of replication and selection is far from obvious, to life forms where these processes are central to evolution. Below I discuss mathematical models combined with experimental evidence that might offer glimpses into how this pivotal transition occurred.

## Origin of life as a reproducer-replicator symbiosis

Replication of genetic information is a (or, perhaps, the) key feature of evolving biological entities, both cellular life forms and MGE. All these entities are endowed with genomes that, in a more abstract language of evolutionary theory, are denoted replicators (2, 39, 40).A replicator can be defined as an information carrier whose propagation depends solely on the information it encodes and does not require physical continuity of its components. All known replicators are digital devices (nucleic acids) that encode information in the sequence of symbols (nucleobases). Obviously, however, life is not limited to information transmission. A sufficient supply of energy and chemical building blocks (nucleotides, in particular), which requires spatial compartmentalization, is essential for the evolution of replicators.

Hence the fundamental split of all propagating biological entities into reproducers, analogue devices that require physical continuity for propagation, and the digital replicators (2). Cells are reproducers with both mutualistic (genomes) and parasitic (MGE) replicators contained inside them, and cell propagation is not limited to genome replication but rather, involves reproduction of the entire cellular organization that provides the niche for the replicators: *Omnis cellula e cellula*. The entire history of life is an uninterrupted, physically continuous tree of dividing cells. In sharp contrast, all diverse genetic elements (GE) including both genomes of cellular life forms and MGE are pure replicators that recruit molecular machinery for some of the functions required for their replication (41).

The extant reproducers (cells) that all host at least one, mutualistic replicator (the genome) are compartments bounded by phospholipid membranes carrying a broad variety of transporters and other transmembrane proteins (42). Compartmentalization is essential for preventing diffusion of small molecules into the environment, keeping their intracellular concentrations at levels sufficient to sustain metabolism as well as genome replication and expression, and also to maintain the integrity of selectable units that consist of the reproducers together with the resident replicators. Cellular membranes perform essential transport functions, that is, selectively import molecules and ions required for cell homeostasis, metabolism and replication, and expel toxic molecules and ions. In respiring cells, membranes channel the energy released in oxidation reactions to produce ion gradients that are then transformed into the energy of the phosphodiester bond of ATP. Although in one class of replicators, viruses, many form membrane-containing virions, viral membranes do not perform transport and energy-producing functions, and viruses, like all replicators, fully depend on cells for energy and building blocks required for replication (43).

In their relationships with the host reproducer, replicators span the entire range from (nearly) full cooperation and a mutualistic relationship with the host reproducer in the case of cellular genomes to commensalism in the case of plasmids, transposons and some temperate viruses, to aggressive parasitism of lytic viruses (44, 45). To the best of our current knowledge, only a mutualistic union of a host reproducer with a resident replicator(s), the genome that carries instructions for the reproduction of the host, qualifies as a life form (organism) because only such a collective unit can provide for the key feature of life, temporally continuous, robust reproduction and the ensuing evolutionary autonomy (46-48). Thus, although the 'point' of life origin perhaps cannot be defined objectively and unambiguously, it appears reasonable, at least, operationally, to posit that life took off when replicators evolved to encode components of the host reproducers, providing for the long-term persistence and evolution of the composite units. Thus, the origin of life can be equated with the MTE in which a new type of collective units emerged, a union of reproducers and replicators

Origin of life often has been discussed in terms of 'metabolism first' vs 'information first' scenarios (49-51), which can be restated as 'reproducers first or replicators first?' The dilemma seems formidable and resembles a chicken and egg problem, given that, in all extant life forms, these two fundamentally distinct types of biological entities are inextricably linked. To the best of my knowledge, there are no 'pure' reproducers in extant life forms: all cells capable of division as well as some organelles, such as mitochondria and chloroplasts, represent an obligatory, mutualistic union of a reproducer and a replicator(s). The essentiality of this union for modern life notwithstanding, when it comes to the origin of life, the dilemma actually appears far-fetched because the emergence of replication is inconceivable without a steady supply of energy and building blocks, which requires proto-metabolism with sufficient temporal stability, that only can be provided by some form of primordial reproducers (52, 53). Thus, realistically, pre-biological evolution must have started with reproducers that initially did not harbor any replicators, but rather encompassed compartmentalized, self-sustaining proto-metabolic circuits (54). The nature of the pre-biological compartments remains a matter of active discussion. Membrane vesicles are strong candidates, and an increasing number of relevant chemical reactions has been demonstrated to occur efficiently within compartments (55-58). However, networks of inorganic compartments are also seriously considered as plausible cradles for life (59, 60). Regardless of the specifics of the primordial chemistry, the principal capacity of the proto-metabolic networks would have been simultaneous production and/or accumulation of nucleotides and amino acids; nucleobases, sugars and simpler amino acids, at least, are readily synthesized abiotically, under a variety of conditions (61-65). Nucleotides, antecedents of modern coenzymes, as well as amino acids, peptides and metal clusters, could catalyze some of the reactions in proto-metabolic networks. ATP would serve as the universal, convertible energy currency already at this stage. The source of energy for the primordial reproducers is a major problem without an unequivocal solution. The primordial membranes are unlikely to have been capable of chemo-osmotic coupling that provides for the ATP synthesis on membranes of modern cells (66). Thus, the primordial reproducers most likely produced ATP via substrate-level phosphorylation.

If and when sufficiently high concentrations of nucleotides and amino acids were reached within the primordial metabolizing compartments, synthesis of oligonucleotides and oligopeptides at non-negligible rates could have become possible. Oligonucleotides can be efficient catalysts of various reactions, that is, the first, simple ribozymes. A compelling case in point are tiny self-aminoacylating ribozymes that can be as small as pentanucleotides but, strikingly, catalyze self-aminoacylation almost as efficiently as protein aminoacyl-tRNA synthetases (67, 68). The catalytic capacity of ribozymes is sequence-dependent, therefore, fixation and amplification of catalytically efficient ribozymes could have been the principal driver of the evolution of replicators. Obviously, emergence of an efficient replicase is indispensable for the evolution of replicators. For many years, although substantial experimental efforts in this direction led to the demonstration of templated synthesis and ligation of oligonucleotides catalyzed by ribozymes (69-72), compact, efficient, processive ribozyme polymerases have remained elusive. However, in a recent breakthrough, a processive RNA-dependent RNA polymerase consisting of only 45 nucleotides was evolved in vitro (73). Another crucial ribozyme activity is peptidyltransferase, the likely ancestor of the ribosome. A peptidyltransferase activity has been demonstrated for a 64 nt ribozyme (74-76). These findings are making the RNA world an increasingly realistic scenario for the evolution of primordial replicators, most likely, relatively short polynucleotides.

From the onset of replicator evolution, two levels of selection would emerge (Figure 2). First, because the rate of RNA replication, like those of other ribozyme-catalyzed reactions, is sequence-dependent, there would be selection for maximization of the replication rate. The replicators evolving solely under the pressure of selection at this level would become parasites of their reproducer hosts because replication incurs cost by consuming resources. However, some replicators endowed with ribozyme activity could become mutualists if they catalyzed reactions beneficial to the reproducer and the benefit exceeded the cost of replication. At this stage, the mutualistic relationship between reproducers and replicators likely would be facultative rather than essential for either partner, but nevertheless, a new, composite unit of selection would emerge, a reproducer-replicator union.

If pre-biological evolution started with reproducers that harbored proto-metabolic networks, some form of reproduction of the compartments encasing the reacting molecules, obviously, was essential for the long-term persistence of such reproducers. Growth and division of lipid vesicles strikingly resembling the division of wall-less bacteria, such as L-forms, has been demonstrated (77, 78). This is a simple process operating on basic physical principles, with vesicles becoming unstable after reaching a critical size and dividing, not requiring complex molecular machineries that are involved in cell division in modern cells (79, 80). In evolutionary biology, selection is by default associated with replication of digital information carriers (nucleic acids). However, primordial reproducers might have been already subject to a primitive form of selection. Evidently, reproduction of proto-metabolizing compartments would not resemble the high precision division of modern cells that is coupled to genome replication and would instead involve stochastic distribution of components among the daughter vesicles (Figure 2). With this type of reproduction, random drift would play a key role in evolution, and the evolutionary process would resemble the stochastic corrector model of primordial evolution (81, 82). Notwithstanding the major contribution of drift, natural selection would already affect the evolution of pure reproducers, through the survival of the fittest, that is, the most stable, temporally persistent compartments (83).

**A model for the origin of life as a reproducer-replicator union**

Analysis of an agent-based mathematical model of the co-evolution of reproducers and replicators revealed some of the conditions that are essential for replicator-reproducer units to outcompete reproducers lacking replicators, thus precipitating the first MTE (54). In the model, evolution of pure reproducers depends on the trade-off between the metabolic cost of division and the probability of successful reproduction of the protocells. Division of protocells with random allocation of resources (hereafter random division, for brevity) and stochastic protocell death shift the trade-off towards advantages of higher successful reproduction probability, despite the increasing metabolic cost. The presence of replicators (GE) in the protocells can have a major impact on the metabolism (47, 84) but, for simplicity, it can be assumed that the positive feedback of the GE is manifested by the increased probability of successful reproduction of the protocell, without explicitly considering metabolism. The trade-off between the successful reproduction probability and the metabolic cost enables the evolution of mutualism between the protocells and GE, whereby mutualistic GE increase the fidelity of protocell reproduction, while the protocells provide resources for GE replication which inevitably, increases the metabolic cost.

As pointed out above, in reproducer-replicator systems, competition and selection would occur at two levels: between replicators (GE) within a protocell, and between protocells carrying different complements of replicators. As in any population of evolving GE, replicative parasites (cheaters) would inevitably evolve concomitantly with autonomous replicators (cooperators) (85, 86). Thus, there would be 4 classes of replicators within protocells: 1) GE encoding their own replication machinery and thus capable of both autonomous replication and supporting the replication of other GE, and beneficial to the protocell (autonomous mutualists), 2) GE depending on autonomous replicators for replication and beneficial to the protocell (non-autonomous mutualists), 3) GE encoding their own replication machinery and thus capable of both autonomous replication and supporting the replication of other GE but useless to the protocell and incurring cost on the latter (autonomous parasites), 4) GE depending on other GE for replication and useless to the protocell, thus incurring costs both on autonomous replicators and the protocell (non-autonomous parasites). The interactions between these distinct types of replicators and between different types of replicators and protocells (reproducers) would shape the dynamics of pre-biological evolution.

Below I explore the setting most relevant for the origin of life, when the interactions among GE occur inside protocells and there is feedback between protocell reproduction and GE replication. Because replication of GE increases the housekeeping cost for the host protocells, protocells that harbor GE can win the competition against those lacking GE only when at least some of the resident GE confer benefits onto the protocells. In the model, the presence of mutualistic GE in a protocell increases the probability of successful reproduction whereas parasitic GE only incur cost. The interplay between the opposing effects of mutualistic and parasitic GE on the reproduction of protocells defines the evolutionary outcome for the entire reproducer-replicator unit. Consider first the interaction and evolution of only two of the four types of GE, autonomous mutualists and non-autonomous parasites. Simulation of the evolution of protocells containing these two types of GE in competition with GE-less protocells revealed three features of the model as being critical for GE-containing protocells outcompeting the GE-less ones (Figure 3). The first critical parameter is the rate of GE replication compared to the rate of protocell reproduction, that is, the number of GE replication events per protocell division. In an evolving population of GE, the target of selection is the fastest replication. However, for the reproduction of GE-containing protocells, this is a losing strategy, for the simple reason that rampant replication of GE consumes the resources required for protocells reproduction, precluding division. As a result, GE-containing protocells die out, even when some of the GE are mutualists, and the GE-less protocells win the competition. Obviously, the GE population goes extinct with the GE-containing protocells. The dynamics of GE in this scenario includes a burst of parasites that replicate faster than mutualists followed by the collapse of the entire population. However, the outcome of the competition reverses when the GE replication is coupled with the protocell reproduction, that is, replication occurs once per protocell division. In this case, the GE-containing protocells win the competition, and moreover, in the steady state, the surviving protocells contain only mutualistic, autonomous GE (Figure 3). The takeover by the protocells that harbor mutualistic replicators is predicated on a key feature of protocell reproduction, the random distribution of GE between the daughter protocells at division. Indeed, the stochasticity of GE distribution provides for the emergence of protocells carrying only mutualistic GE that win the competition (Figure 3). A similar conclusion has been previously reached in the analysis of another a model of group selection of replicators within cells (87). Modern cells possess tightly orchestrated division mechanisms that ensure symmetric allocation of replicators (copies of the genome) between the daughter cells. Such mechanisms could not exist at early stages of protocell evolution, and actually, would have been counter-productive. In model simulations with symmetric GE distribution at division, the GE-containing protocells went extinct due to the takeover by parasitic replicators (Figure 3). The third key parameter is the metabolic efficiency of the protocell: utilization of the resources must be efficient enough to suffice for both protocell reproduction and GE replication.

Consider now protocells containing replicators of the other two classes, non-autonomous mutualists and autonomous parasites (Figure 4). The probability of successful reproduction of the protocells increases with the fraction of (in this case, non-autonomous) mutualists, hence both protocell-level and GE-level selection favor mutualist-only protocells. However, in this case, such protocells are not sustainable because the mutualistic GE are incapable of replication, and accordingly, all GE will be lost. Protocells containing only autonomous parasites will lose the competition to GE-less protocells because these GE provide no advantage to the protocell and, on the contrary, incur a cost by consuming resources. In this case, to enjoy a sustainable competitive advantage, the GE-containing protocells would have to carry both types of GE. However, such protocells will be outcompeted by protocells containing only non-autonomous mutualists due to the cost of the parasites (Fig. 4). Therefore, with this combination of replicator types, there is no path for GE-containing protocells to take over the population. Extrapolating these findings to protocells containing all 4 types of replicators, takeover by GE-containing protocells is achievable only through the emergence of protocells containing solely autonomous mutualists, regardless of the initial combination of replicator types. The primordial replicators must have been relatively small, most likely, RNA molecules. The modeling results indicate that, to persist in the protocell population, each of these GEs must have encoded their own replication machinery, even if encoding other information beneficial for protocells reproduction, resulting in substantial cost due to resource expenditure. Thus, selection must have favored mergers of these GE into large ones, each encoding a single copy of the replication apparatus. Eventually, the selective pressure for resource economy would result in the formation of large genomes resembling modern prokaryote chromosomes.

**Multilevel selection as the key to the origin of life**

Long-term genetic memory stored on stable digital information carriers – nucleic acid molecules - is an inalienable attribute of life. The persistence of these information carriers (genomes) requires efficient, compartmentalized metabolic networks that provide sufficient amounts of chemical building blocks and energy for genome replication and expression. Thus, any living organism is an amalgam of informational and operational systems, or put another way, a union of a reproducer(s) and a replicator(s). Although the origin of life is often discussed in terms of alternative, metabolism-first or replication-first scenarios (49-51), realistically, this dilemma appears to be false because sustained replication is impossible without the support provided by metabolic networks. Hence the hypothesis on the origin of replicators inside primordial reproducers - protocells. Protocells are often perceived as entities that already contained replicating RNA molecules within lipid membrane-bounded vesicles (88). However, an early stage of prebiological evolution involving selection for persistence (83) among reproducers lacking any replicators (GE) appears to be logically unavoidable. These protocellular reproducers would become hatcheries for primordial replicators that, probably, emerged as parasites but could be fixed in evolution only after some of them became mutualists. At this stage, multilevel selection – that is, selection occurring both at the level of GE and at the level of protocells containing or lacking GE - would take off (Figure 5). Although it is difficult or perhaps impossible to define the point of the origin of life objectively, the onset of multilevel selection seems to be a strong candidate. Thus, at the first and most fundamental MTE – the origin of life itself – the ETI corresponded to the emergence of a new type of Darwinian units, the reproducer-replicator union. This first MTE, then, qualifies as an egalitarian transition because two distinct types of entities merge to form the new, collective unit, but also as an autogenic transition because one of the merging entities (replicators) came from within the other (reproducers) (6).

Replicators present a major, inherent problem because their evolution inevitably and repeatedly gives rise to parasitic GE that hijack the replications machinery of autonomous elements (85, 86). In well mixed systems, the parasites that replicate faster than autonomous elements and therefore take over, leading to the eventual collapse of the entire replicator ensemble (89). Compartmentalization substantially changes the evolutionary dynamics of autonomous and parasitic GE, preventing the takeover by parasites and stabilizing the system as a whole (89-93). Furthermore, previous modeling studies strongly suggest that replicators are more likely to survive within protocell-like compartments than in surface-based spatial systems (94). In the models of reproducer-replicator coevolution, compartmentalization combined with the stochastic allocation of GE at protocell division provides for the formation of protocells containing only autonomous mutualists which is a necessary condition of the fixation of GE in the protocell population. Crucially, the coevolution simulations show that, for the fixation of GE-containing protocells, the GE replication has to be coupled with the reproduction of the protocells. Although, in isolation, evolving GE would be selected for the highest replication rate, this is not the case when GE survival depends on the reproduction of protocells. Indeed, GE replication at rates substantially exceeding the reproduction rate of the protocells leads to the extinction of the GE-containing protocells due to the exhaustion of the resources available for protocell reproduction. Thus, at the first MTE, competition and selection at the lower level, that of individual replicators, would be partially suppressed to enable the persistence of the new, composite Darwinian individual, the GE-containing protocell. In this respect, the first MTE was apparently analogous to the subsequent, better understood ones, such as the emergence of multicellularity where proliferation of individual cells is partially suppressed ensuring the survival of the composite unit, the multicellular organism, whereas uncontrolled cell proliferation leads to collapse of the system (cancer).

The multilevel selection model discussed here suggests an evolutionary path towards large genomes,

comparable to chromosomes of extant bacteria and archaea. There is little doubt that the first GE were small, on the order of a kilobase, at most. As discussed above, GE-containing protocells could win the competition with GE-less ones only when all or at least a large fraction of the GE within a protocell population encoded their own replication machinery. Thus, at the early stages of evolution, the typical GE most likely resembled the simplest of the modern RNA viruses that encode their own replicase (RNA-dependent RNA polymerase) and, in some case, one or two additional proteins (95, 96). The protocells that harbored populations of autonomous, mutualistic GE would encompass multiple versions of the replication machinery. This redundancy and the ensuing extra cost could not be simply dispensed of by deleting the sequences encoding the replication system because loss of the replication machinery would turn the respective GE into parasites, potentially. resulting in the collapse of the GE population. Hence, selective pressure for joining GE and eliminating the redundancy, keeping a single copy of the replication machinery, saving resources and facilitating coordination of GE (genome) replication with protocell division.

The origin of chromosomes was likely coupled with the evolution of cell division. As emphasized above, at the earliest stages of evolution, stochastic allocation of GE at protocell division was apparently essential for the fixation of GE in protocell populations. Once primordial mutualist replicators merged into large genomes present in a single or few copies per (proto)cell, accurate segregation of genomes into the daughter (proto)cells would become advantageous, creating selective pressure for the mechanisms of symmetrical division. However, symmetrical division makes (proto)cells vulnerable to parasite onslaught because parasitic GE would persist after invading or evolving from mutualists by mutation. Therefore, defense mechanisms, conceivably, based on specific recognition of parasite sequences, would coevolve with symmetrical cell division mechanisms, being a pre-requisite for the long-term survival and evolution of cells. Defense systems are enormously abundant and diverse in modern prokaryotes where they account for a substantial, likely, still underestimated fraction of the genome (24).

The symbiotic scenario for the origin of cells presented here is compatible with the RNA world concept (88, 97, 98). The key point, however, is that the primordial RNA world – that is, populations of RNA replicators, some of them endowed with ribozyme activity - must have evolved within pre-existing, metabolically active, membrane-bounded protocells (reproducers) as previously proposed by Copley, Smith and Morowitz (99).

Although deliberately presented in abstract terms, the model of the symbiotic origin of life presented here suggests many avenues for experimental validation. In particular, experimental modeling of the origin of replicators within reproducers, that is, membrane vesicles encompassing proto-metabolic networks producing nucleotides and amino acids, and potentially, oligonucleotides and peptides, might not be far beyond the capability of modern laboratories (100, 101). Indeed, a major step has been made very recently when a synthetic cell with a 90 kb genome spread over 8 plasmids and capable of genome replication, growth, division and selection was created in the laboratory (102). Between such developments in synthetic biology and advances in ribozyme chemistry, testing the symbiotic scenario for the first MTE might be feasible in a foreseeable future.

**Author contributions**

E.V.K. wrote the paper.

**Acknowledgements**

I thank Puri Lopez-Garcia for the initial inspiration on the symbiotic concept of the origin of life and many indispensable discussions. E.V.K. is supported by the Intramural Research Program of the National Institutes of Health (NIH). The contributions of the NIH author(s) are considered Works of the United States Government. The findings and conclusions presented in this paper are those of the author(s) and do not necessarily reflect the views of the NIH or the U.S. Department of Health and Human Services.

**Competing interests**

The author declares no competing interests.

Figures

**Figure 1.** Multilevel selection: organismal, sub-organismal and super-organismal levels**.**

**Figure 2.** Multilevel selection and reproducer-replicator union at the origin of life:

the first major transition in evolution (MTE).

**Figure 3.** Competition between protocells containing and lacking genetic elements: autonomous mutualists and non-autonomous parasites.

**Figure 4.** Competition between protocells containing and lacking genetic elements: non-autonomous mutualists and autonomous parasites.

**Figure 5.** Onset of multilevel selection and the symbiotic scenario for the origin of life.

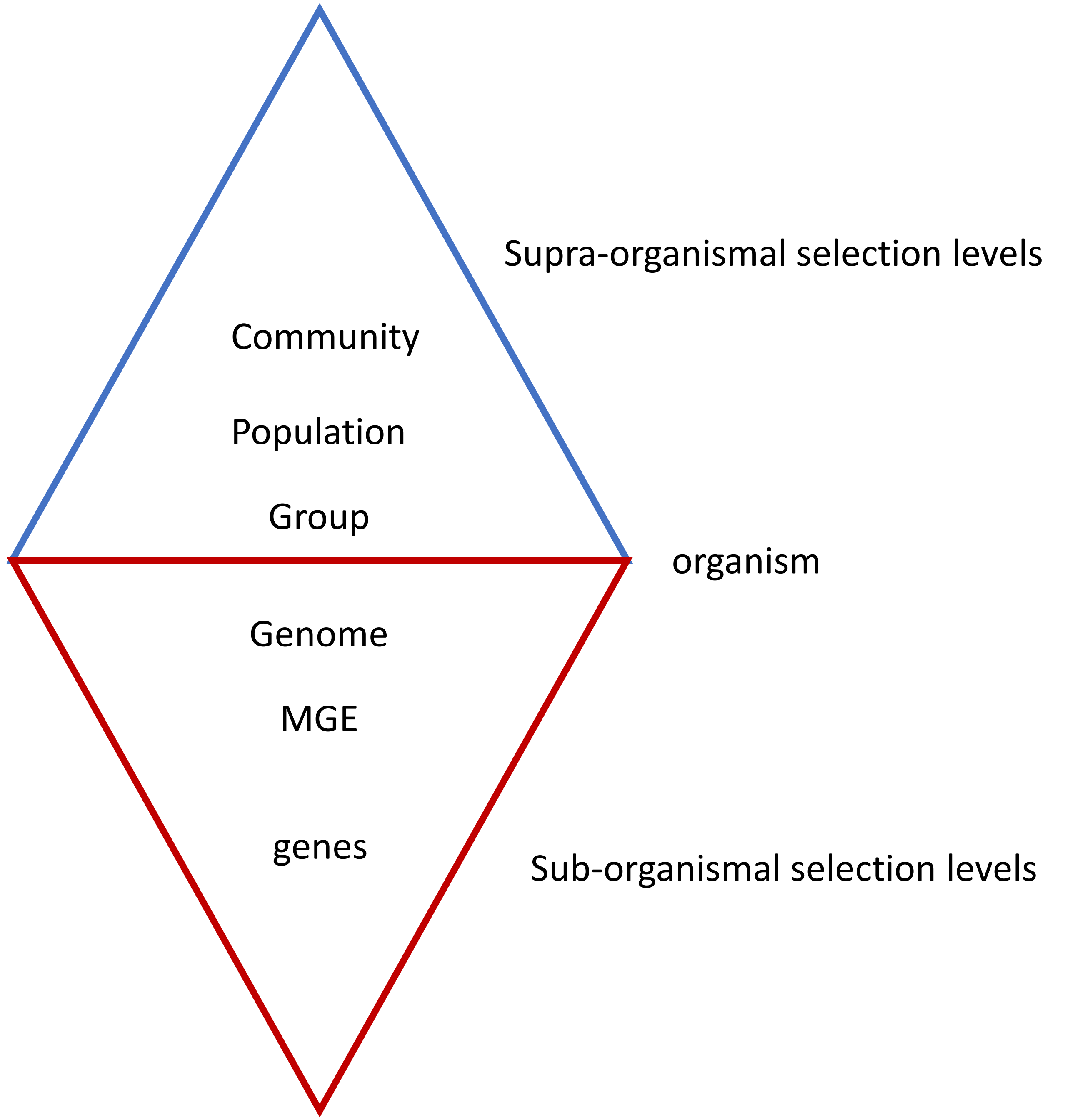
Supra-organismal selection levels
Community
Population
Group
organism
Genome
MGE
genes
Sub-organismal selection levels

Fig. 1

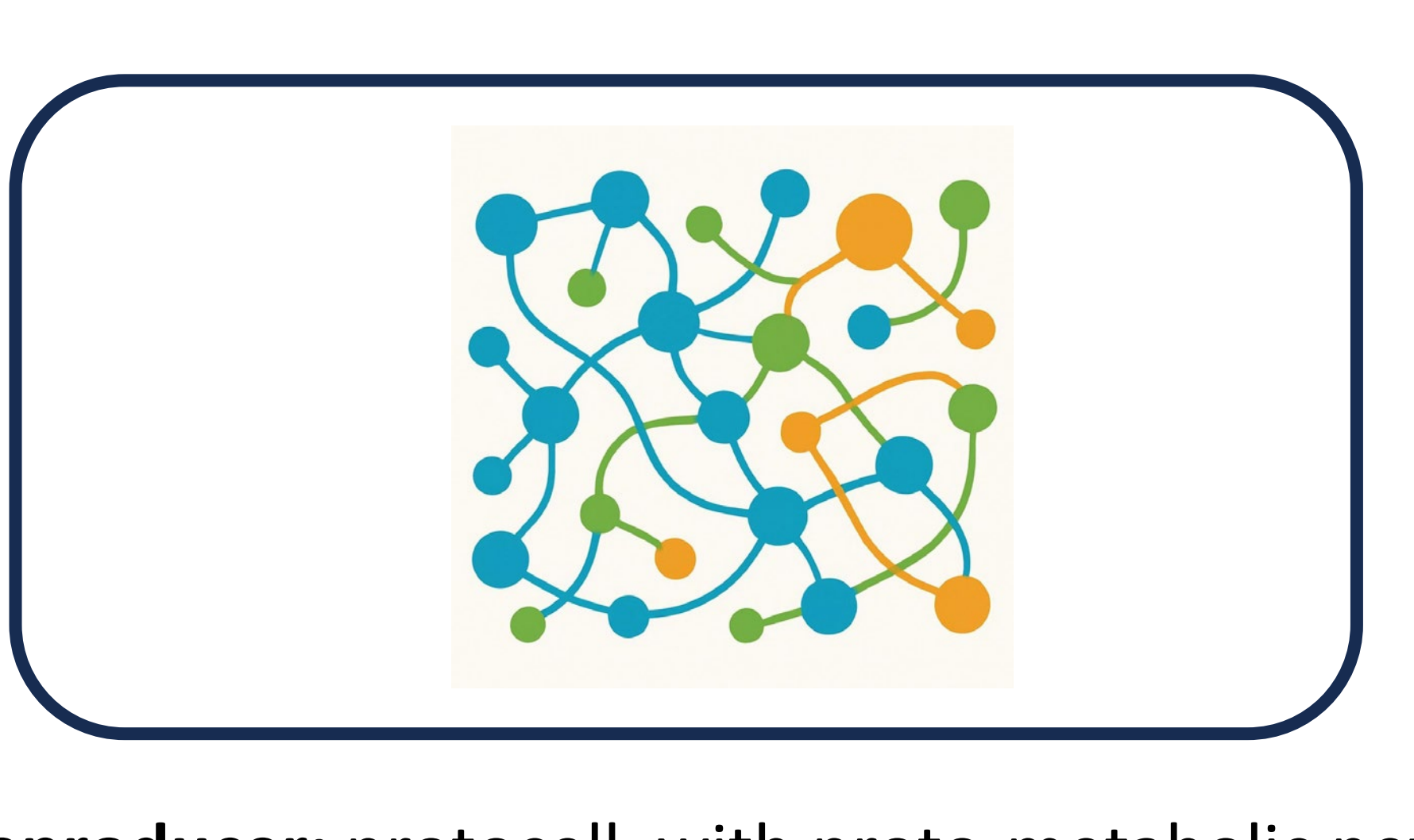

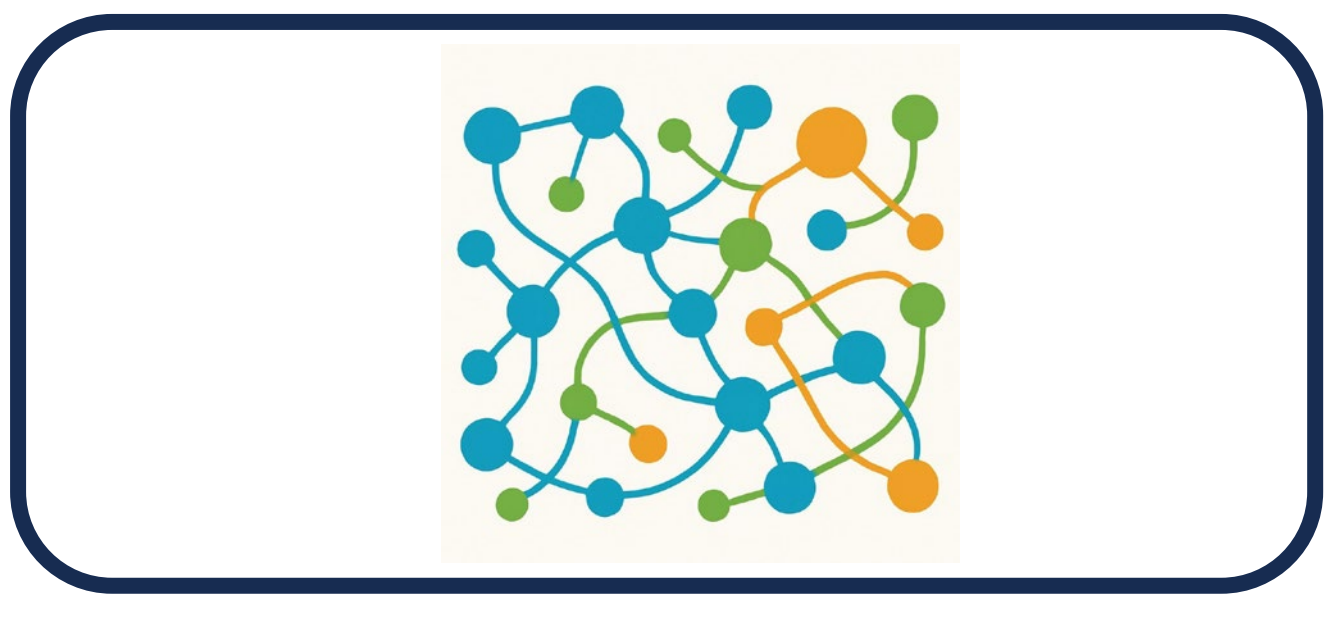

Primitive protocell division

**Selection for protocell Persistence/reproducton**

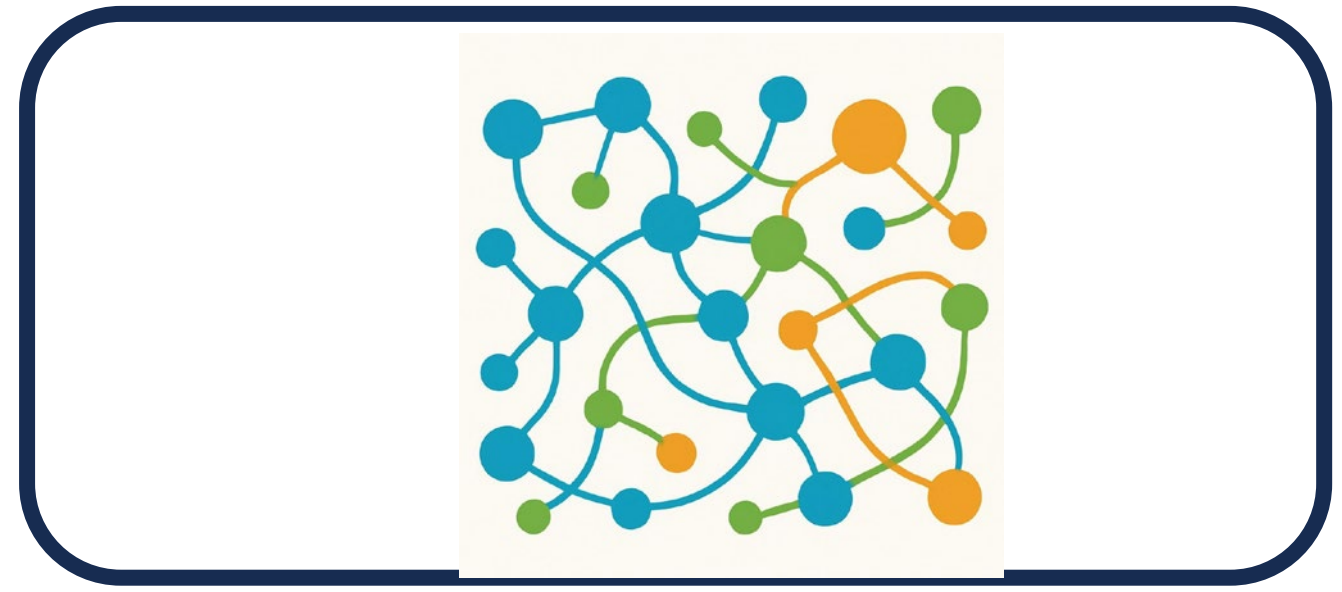

**Reproducer:** protocell with proto-metabolic network

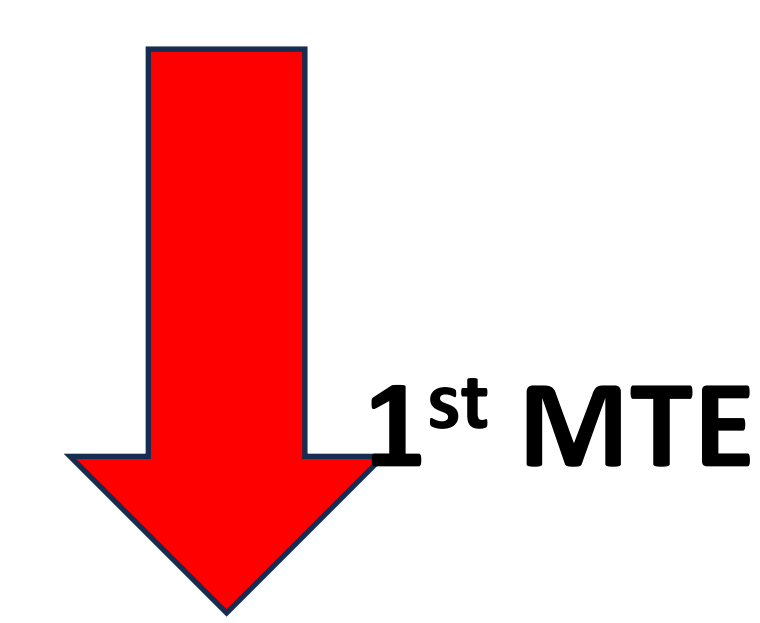

**Replicators:** selection for GE replication rate

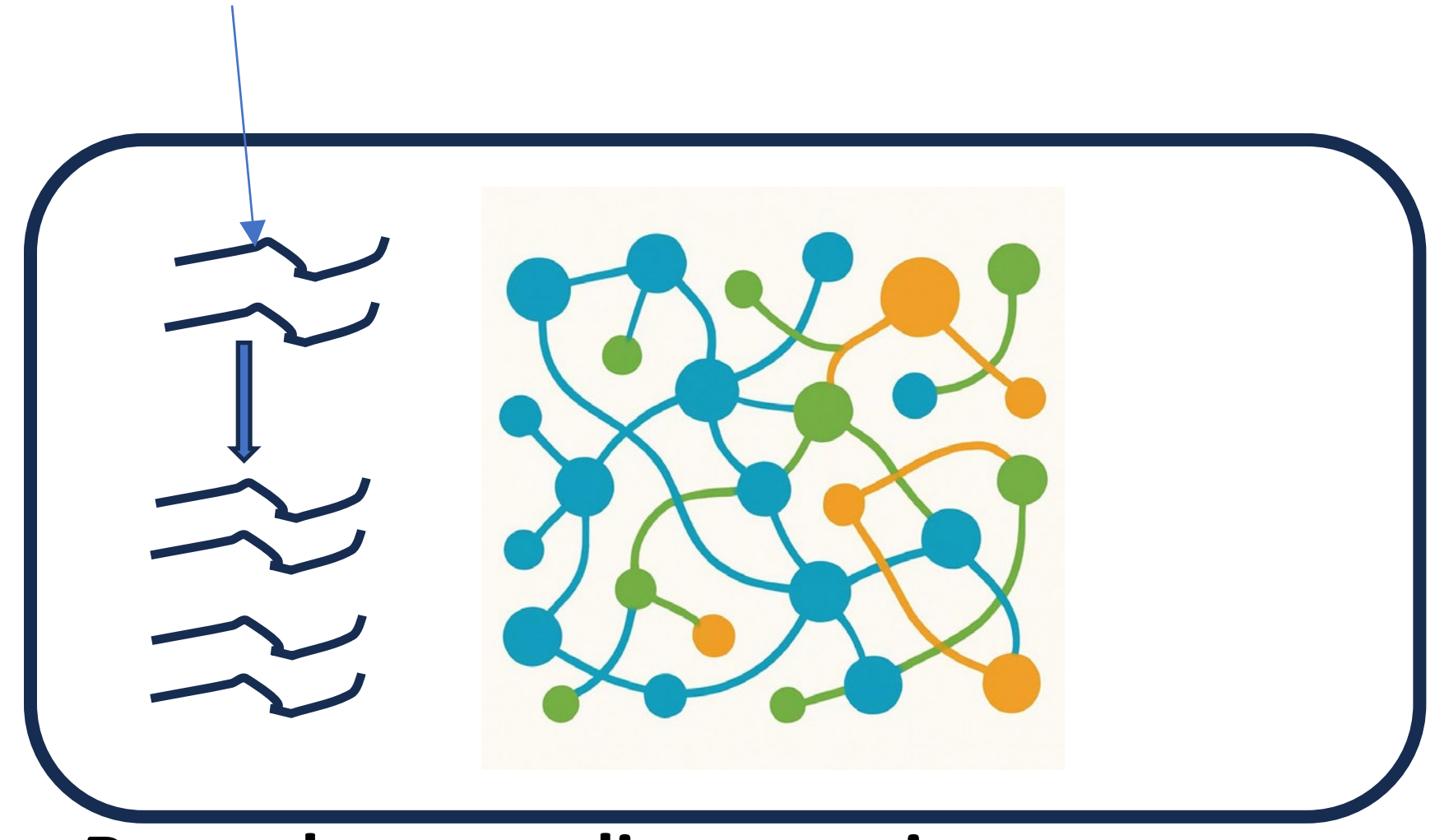

Primitive protocell division with random GE allocation

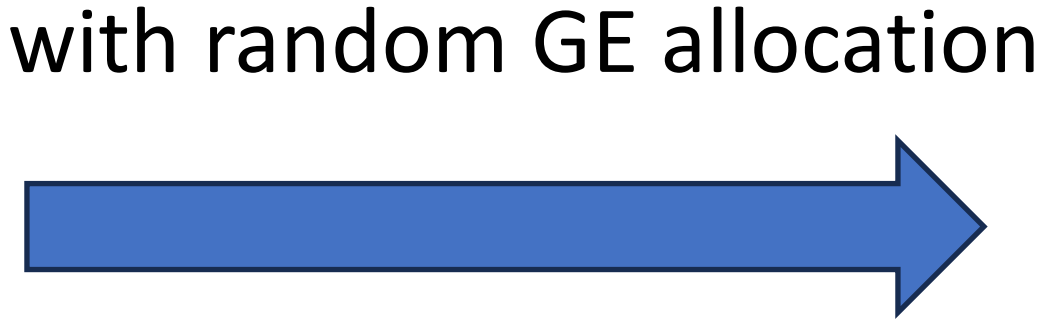

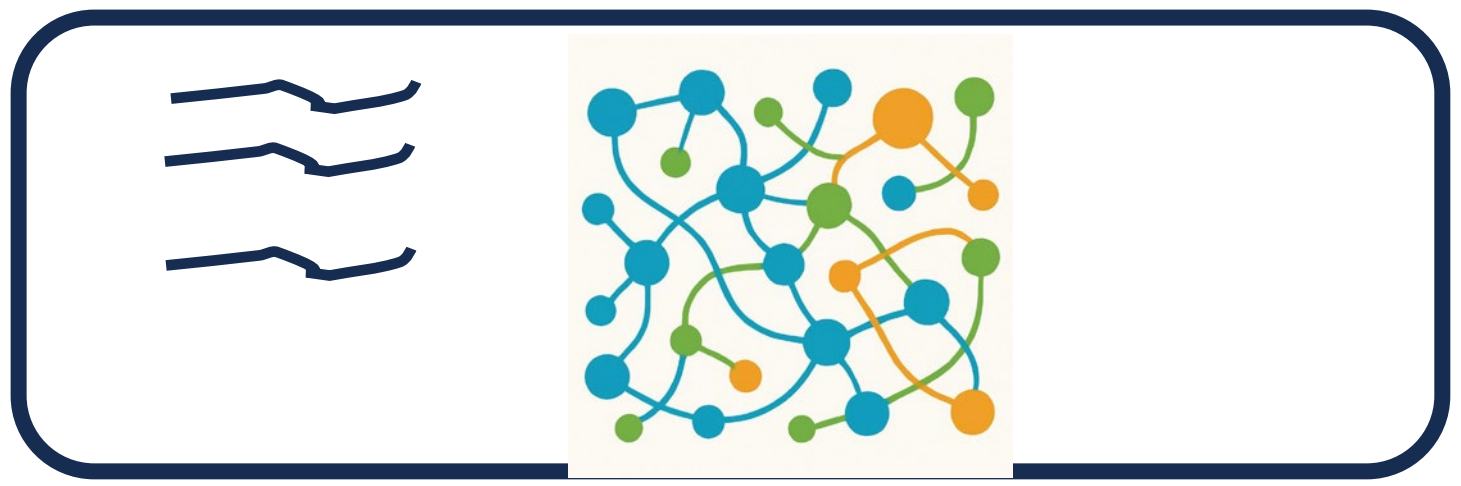

**Selection for reproduction of reproducer-replicator units**

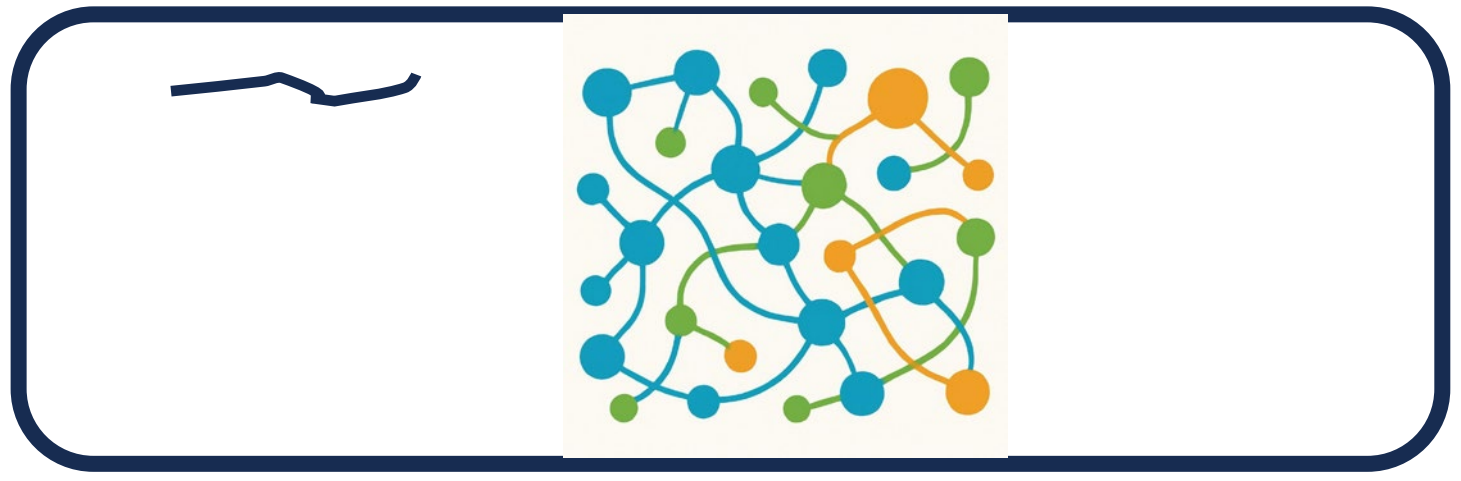

**Reproducer-replicator union**: protocell with metabolic network and GE

Fig. 2

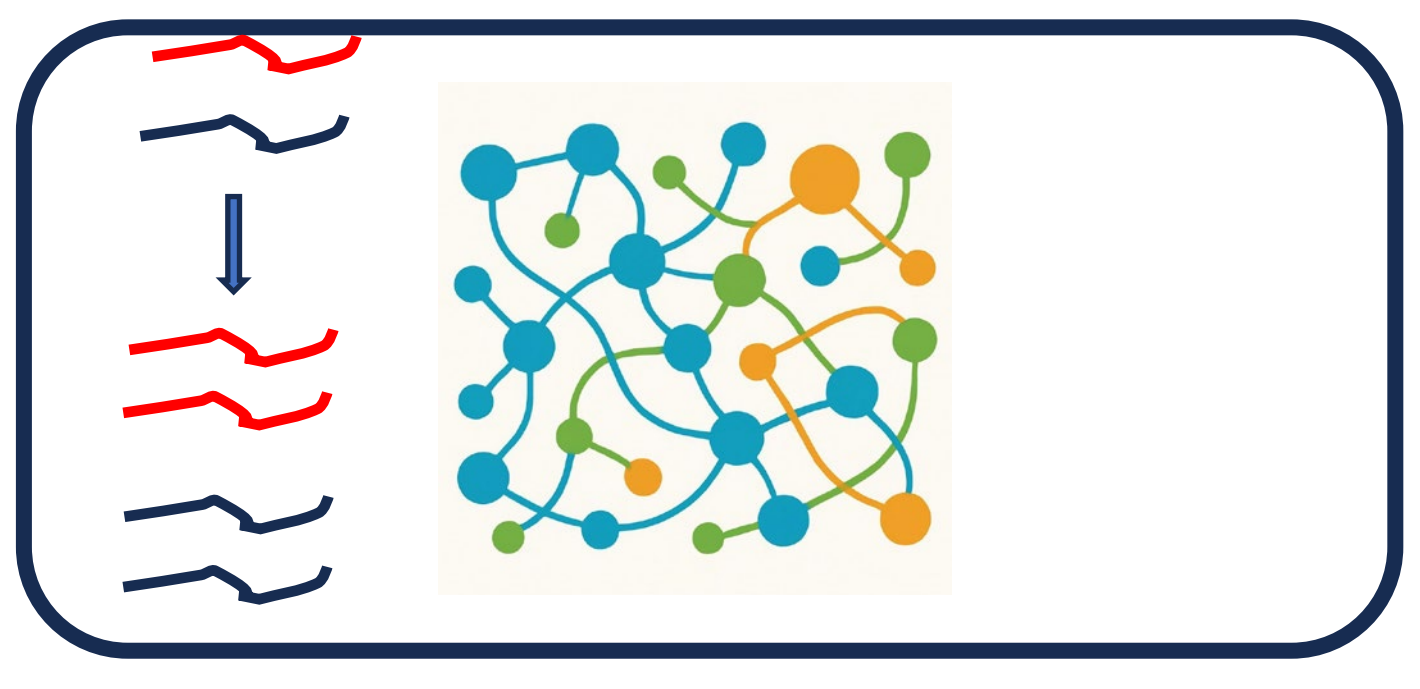
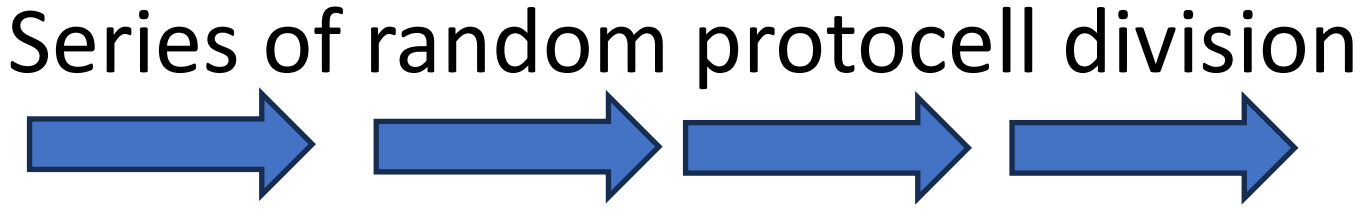
Series of random protocell divisions
Protocells containing only
autonomous mutualists emerge
and win competition
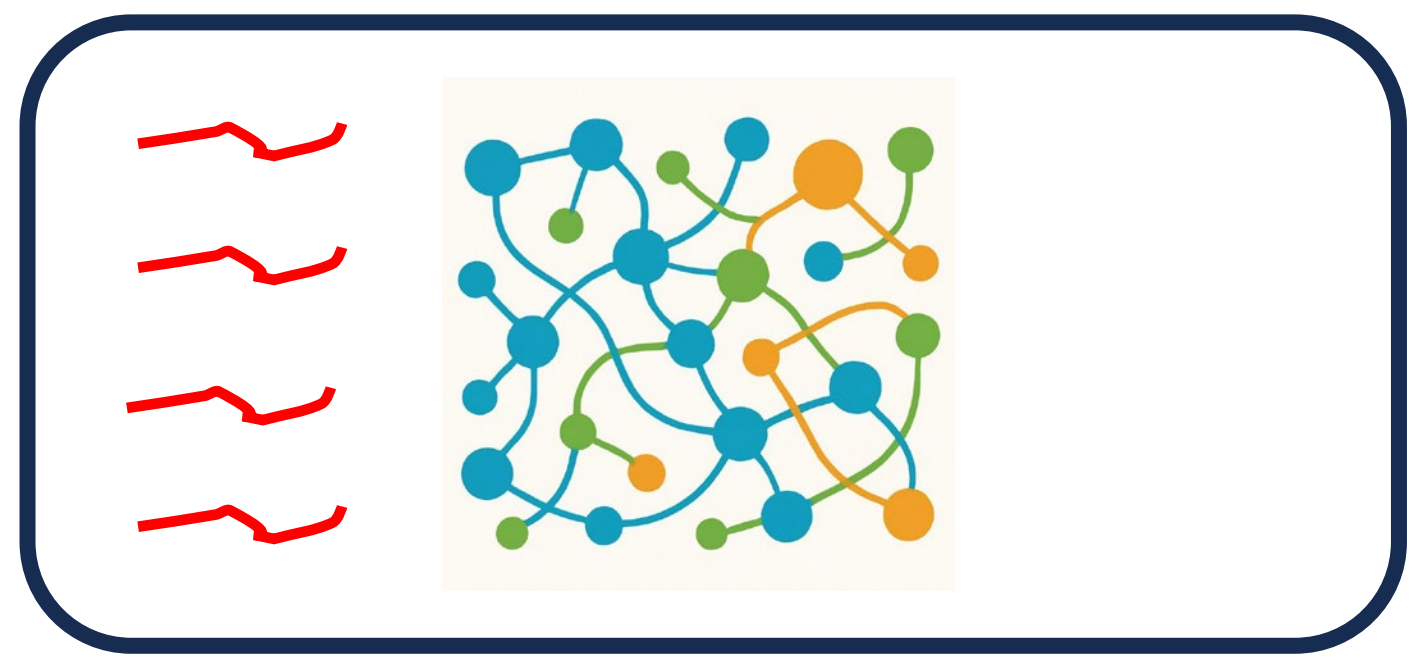
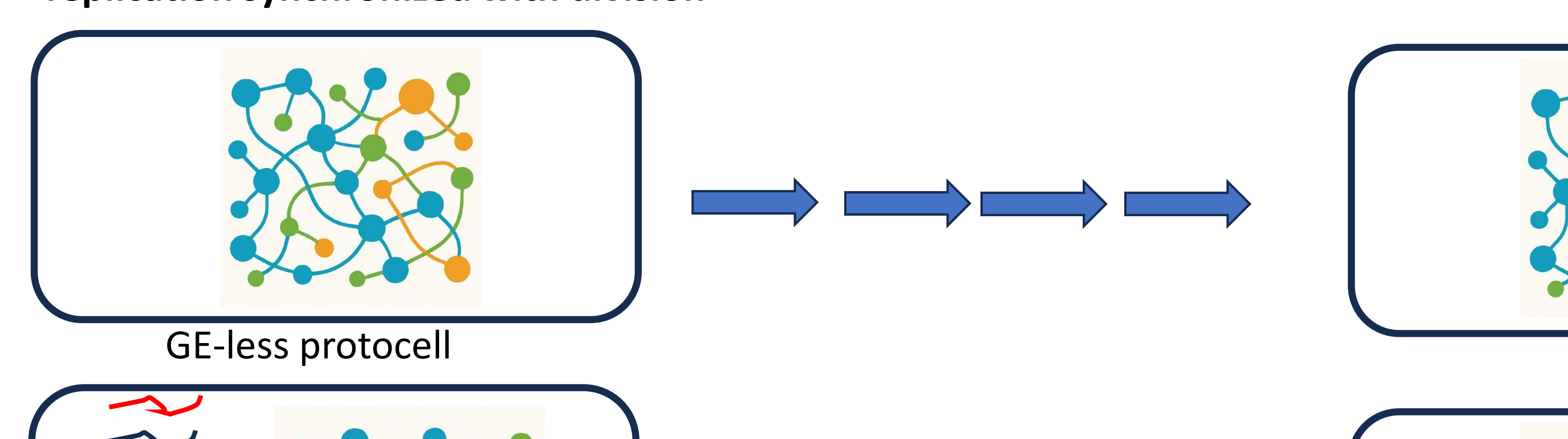
Protocell with autonomous mutualist
and non-autonomous parasite GE,
**replication synchronized with division**
GE-less protocell
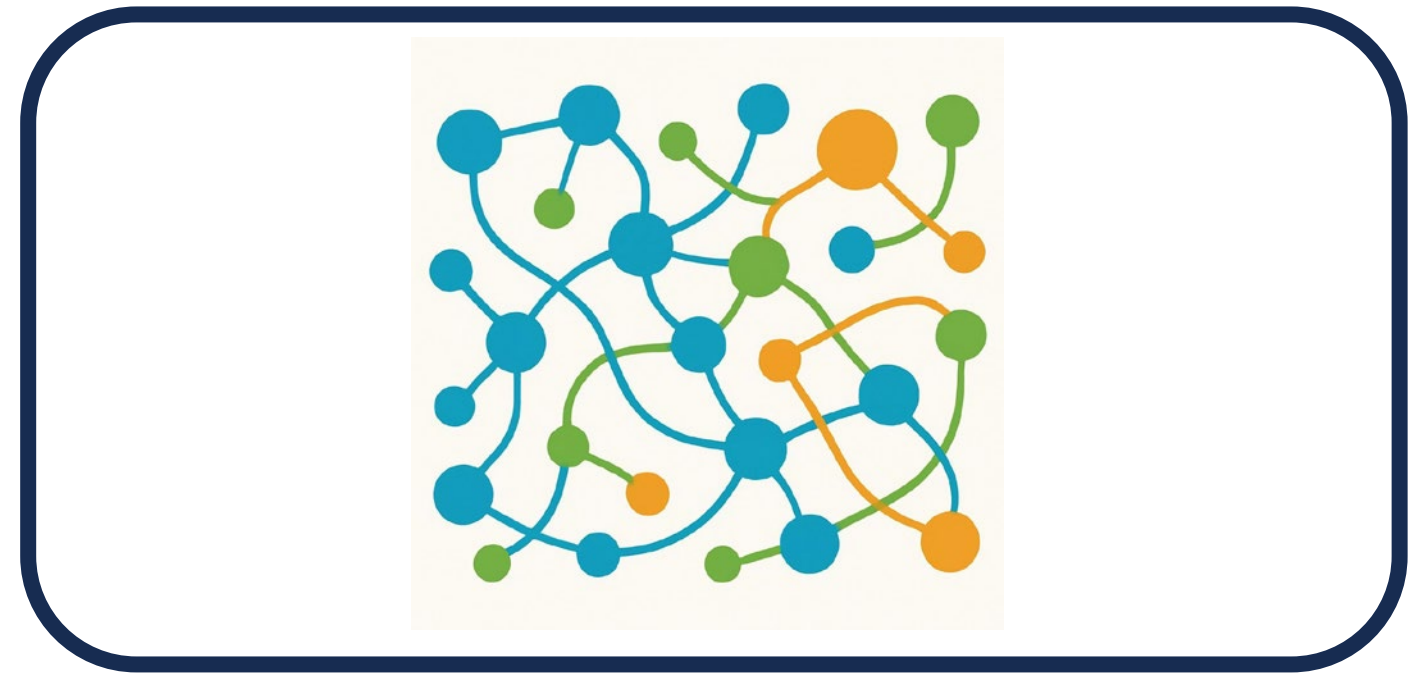
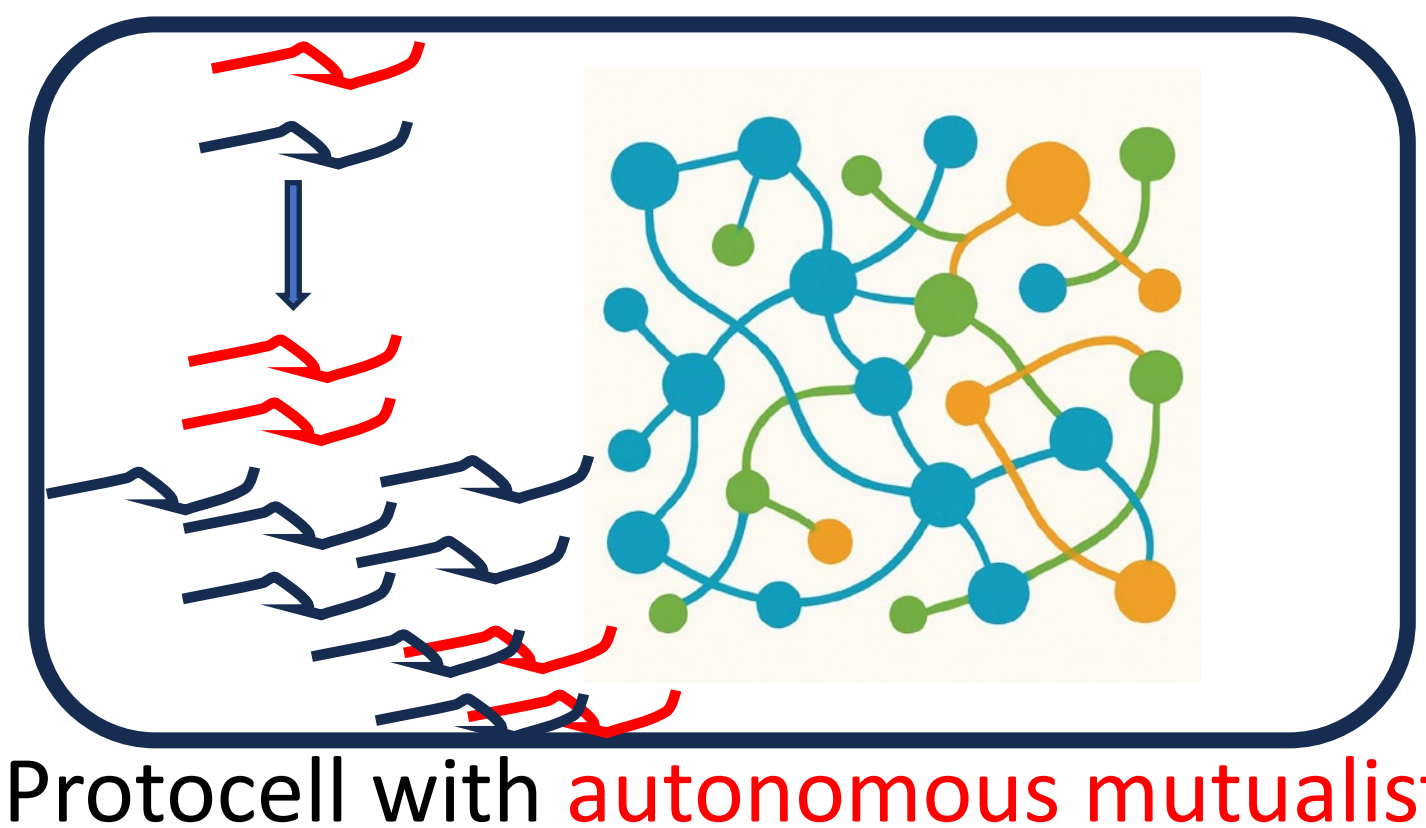
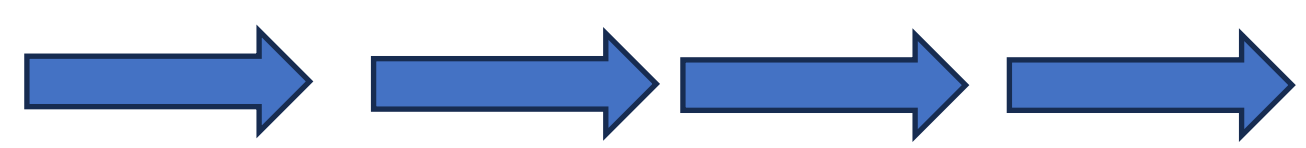
Protocells containing numerous
GE die out or lose GE
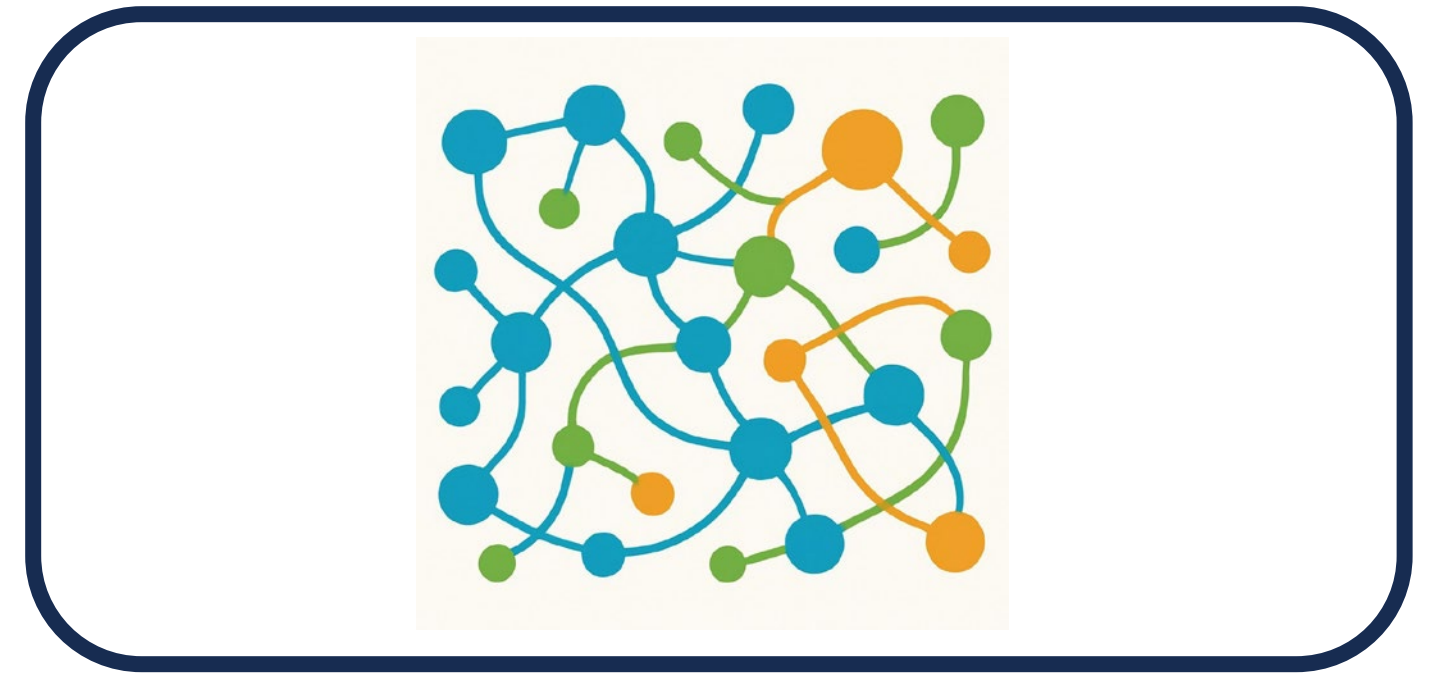
Protocell with autonomous mutualist
and non-autonomous parasite GE, **Rampant replication**

Fig. 3

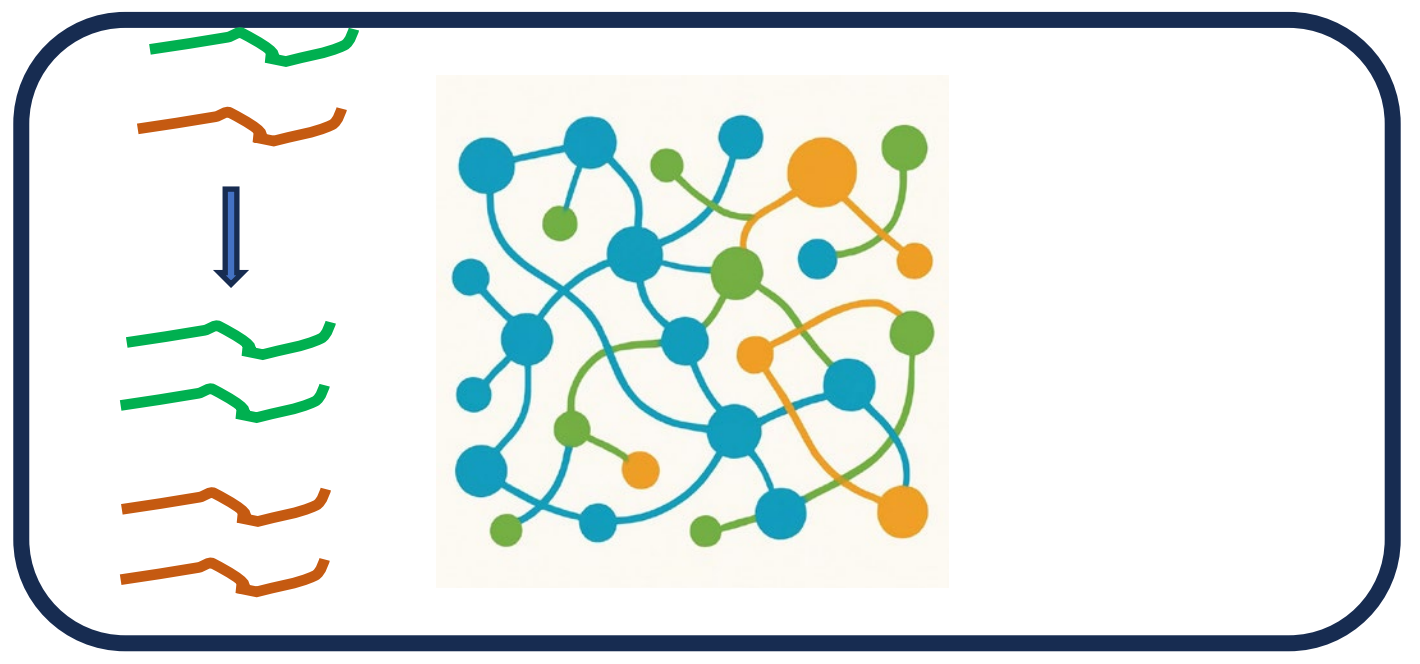

Protocell with non-autonomous mutualist and autonomous parasite GE, replication synchronized with division

Series of random protocell divisions

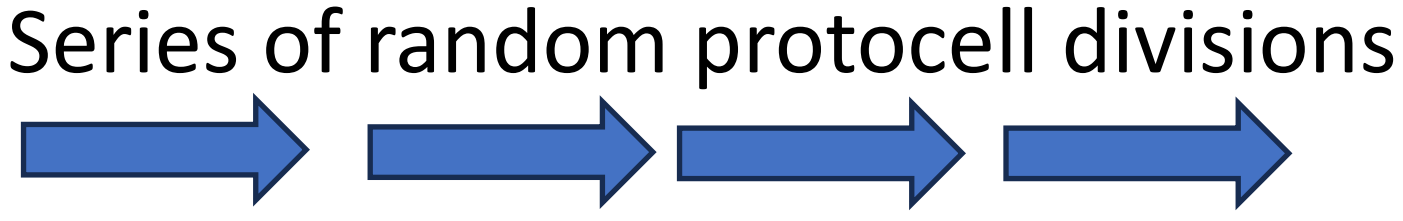

Protocells containing either only non-autonomous mutualists or only autonomous parasites (or a mix) lose competition to GE-less protocells

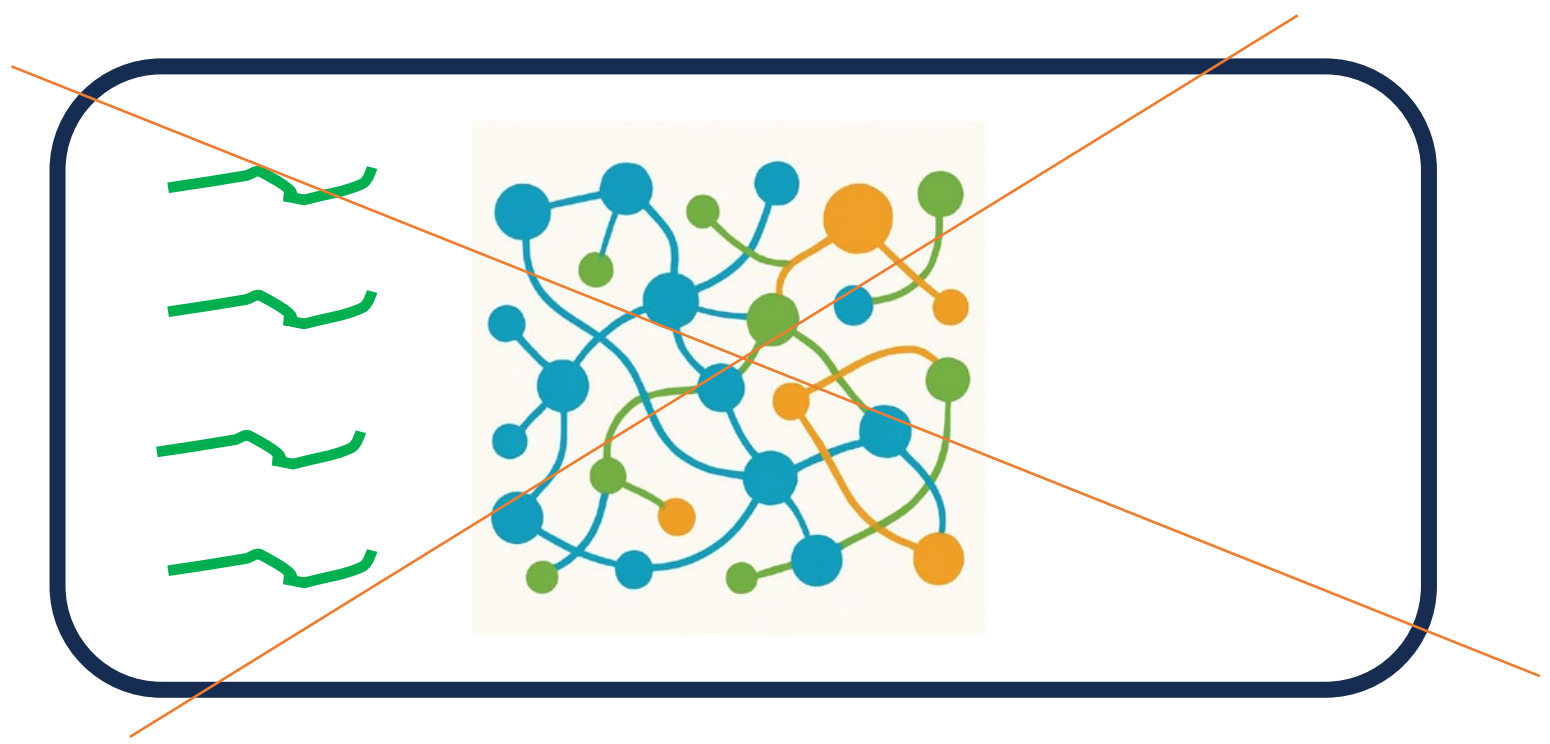

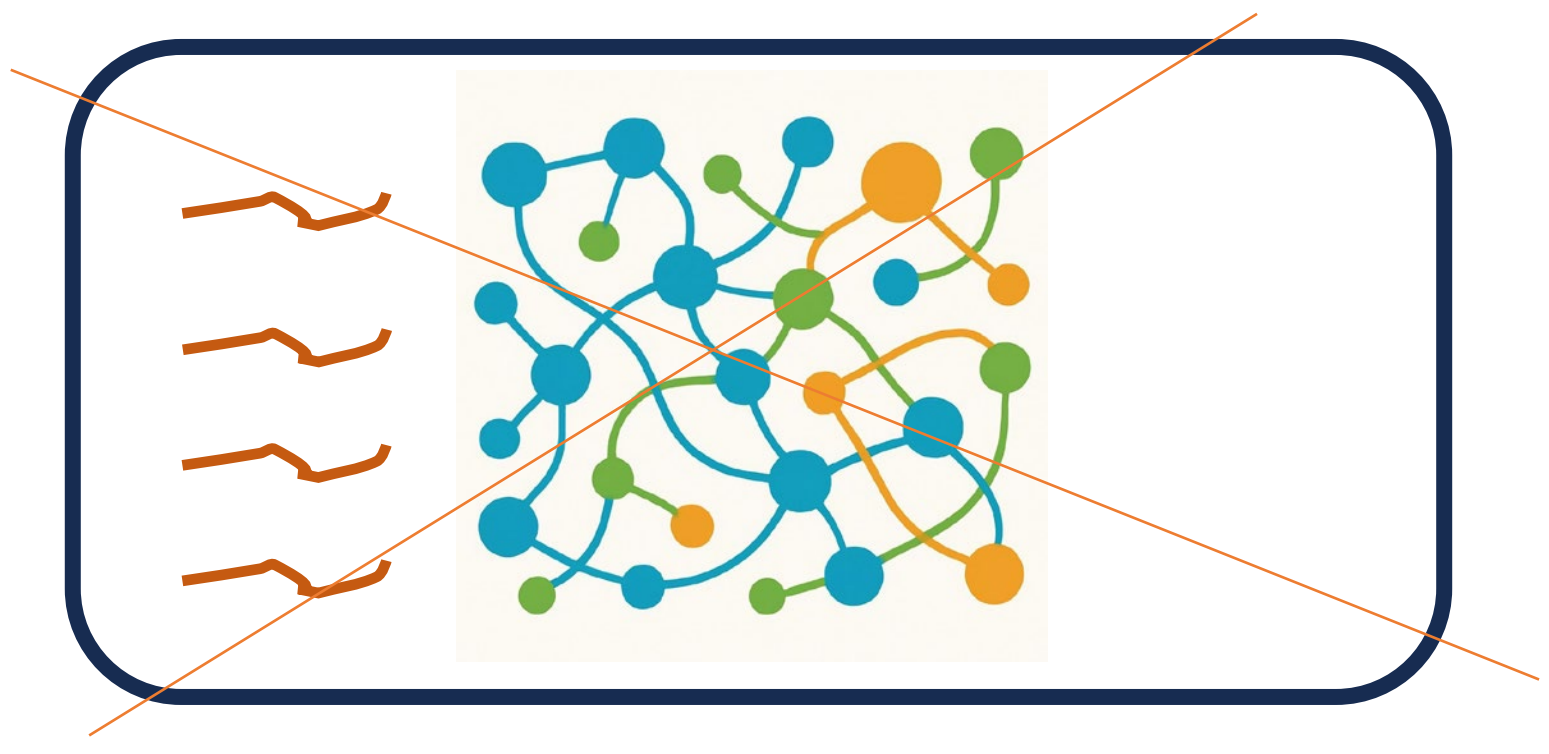

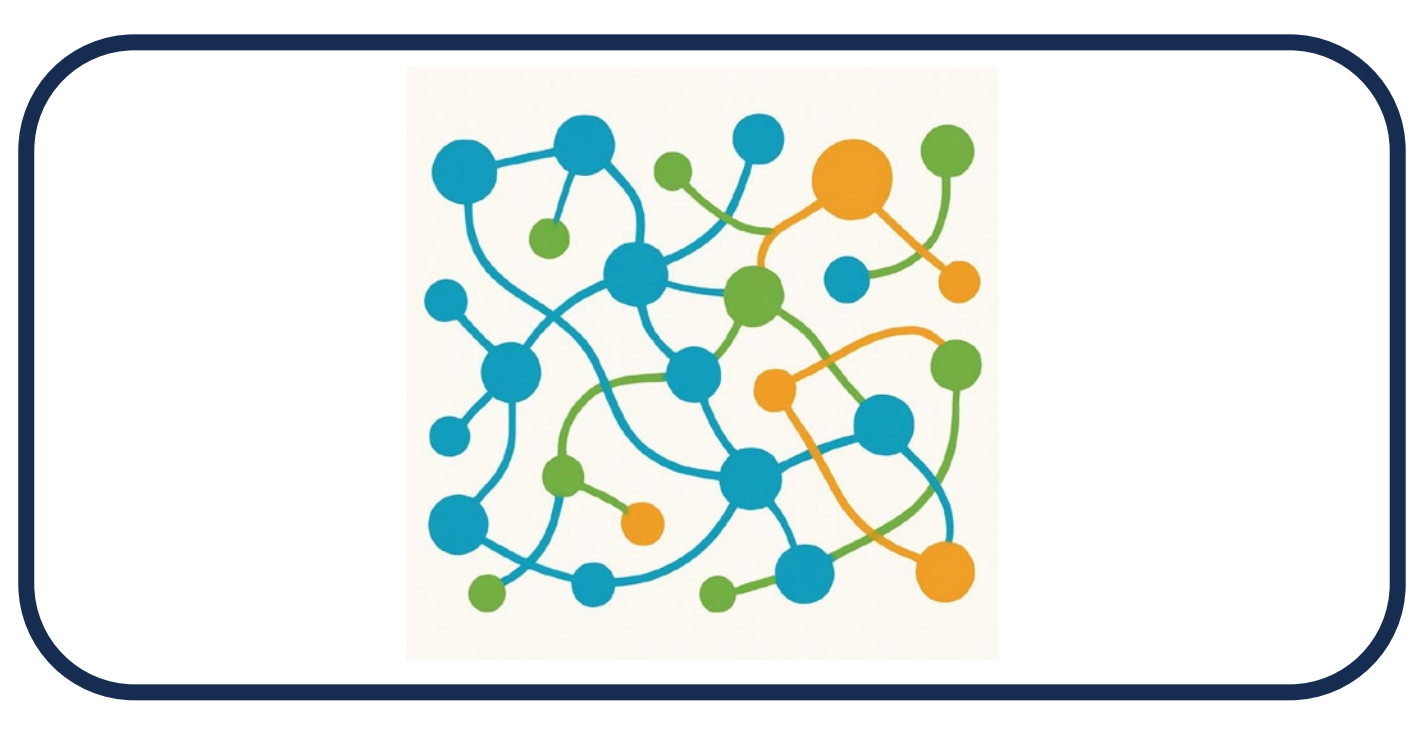

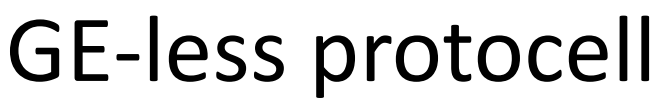

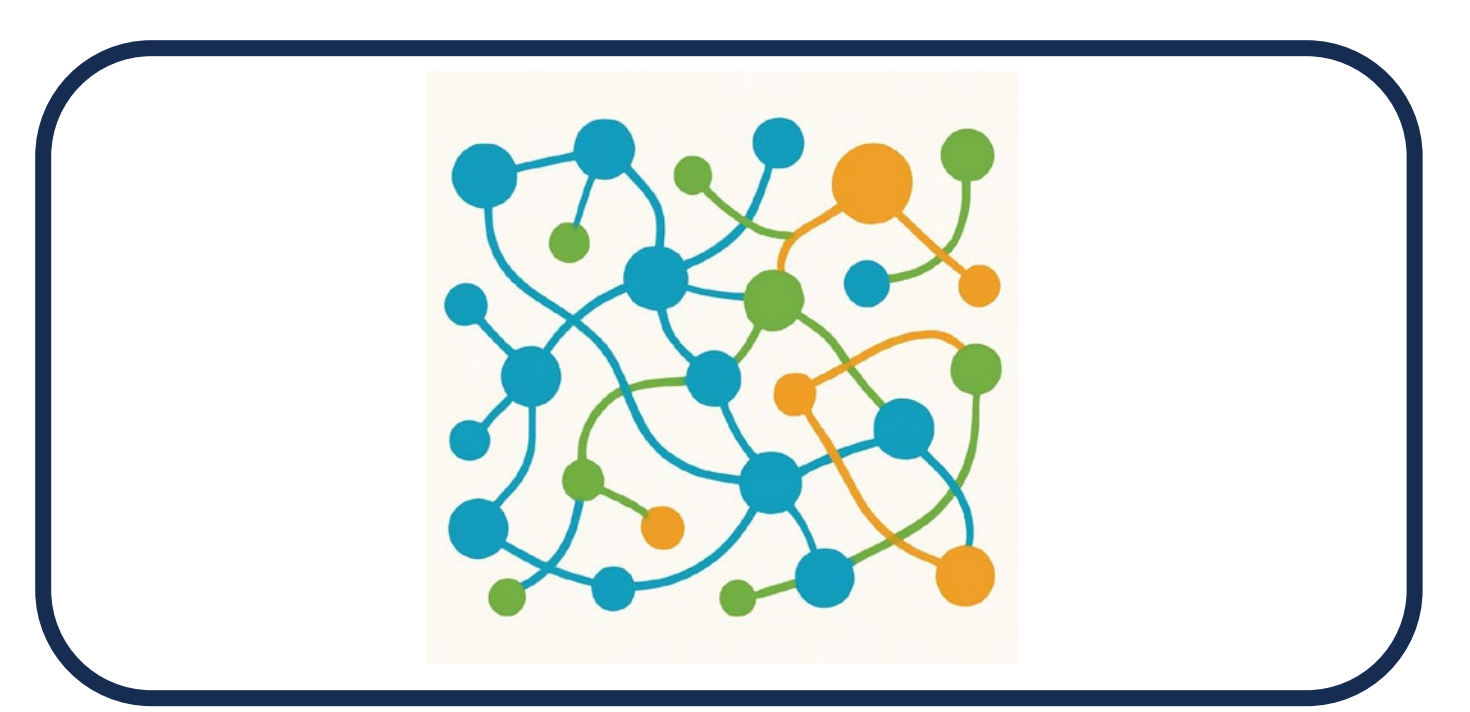

Fig. 4

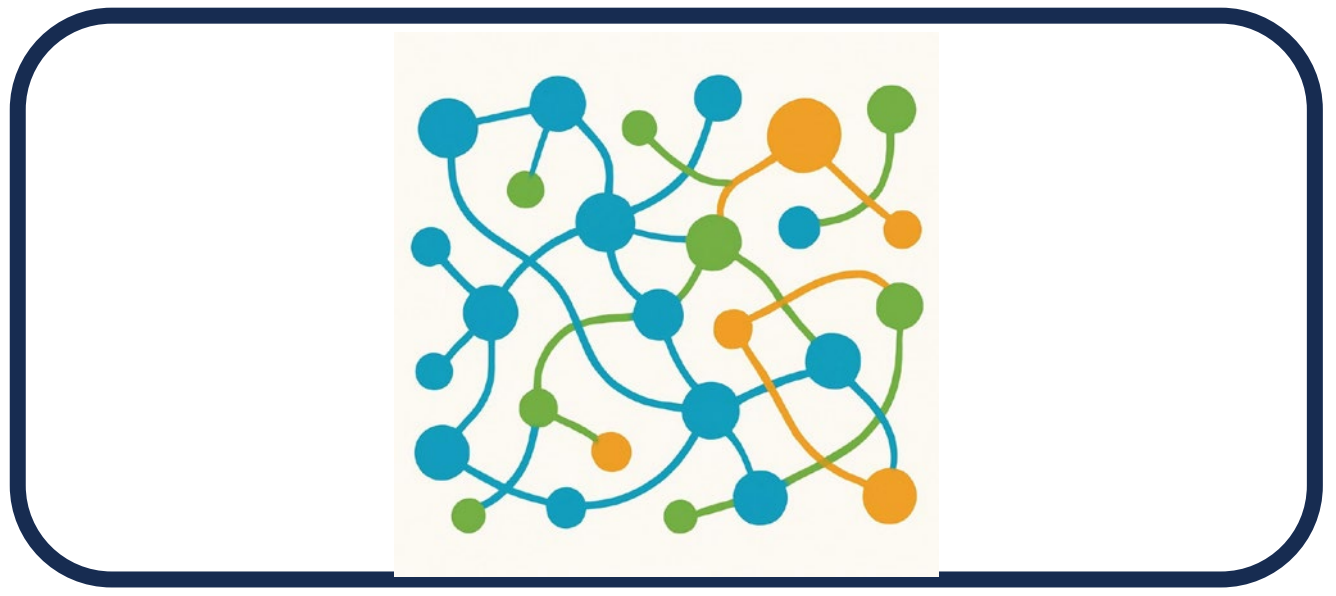

**Reproducer**: protocell with proto-metabolic network

Primitive protocell division

**Selection for protocell persistence**

Selection for replication speed

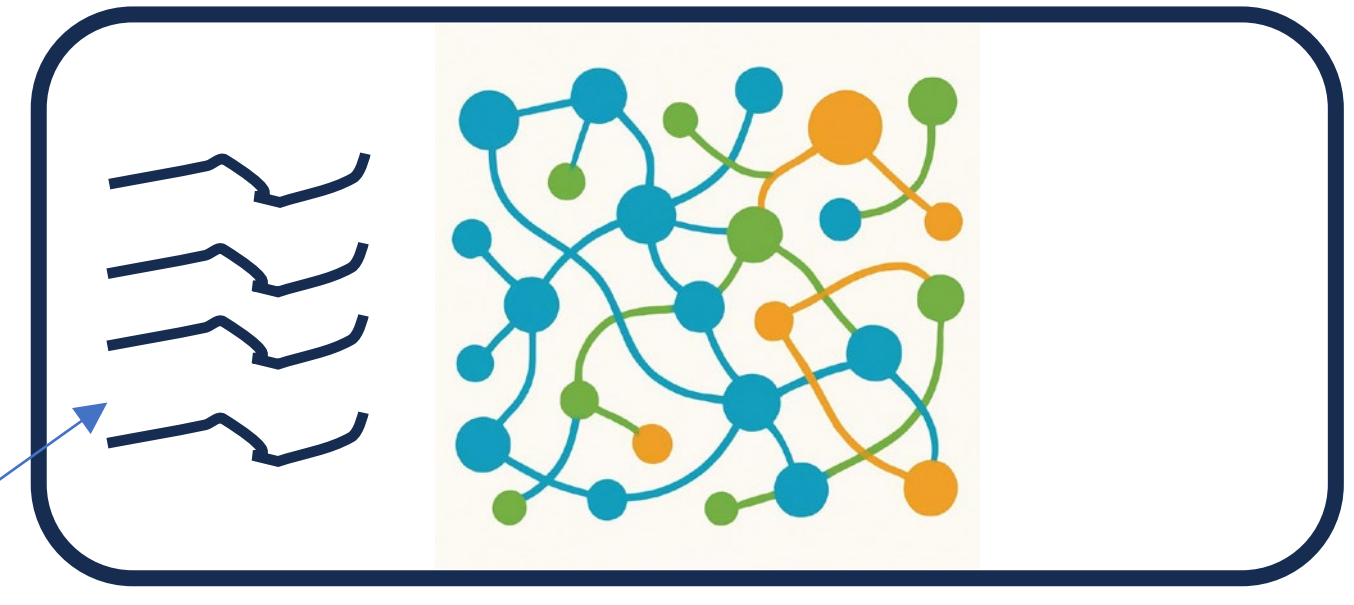

**Reproducer/replicators**: protocell with proto-metabolic network and parasitic replicators

Random division,
Emergence of
mutualist GE

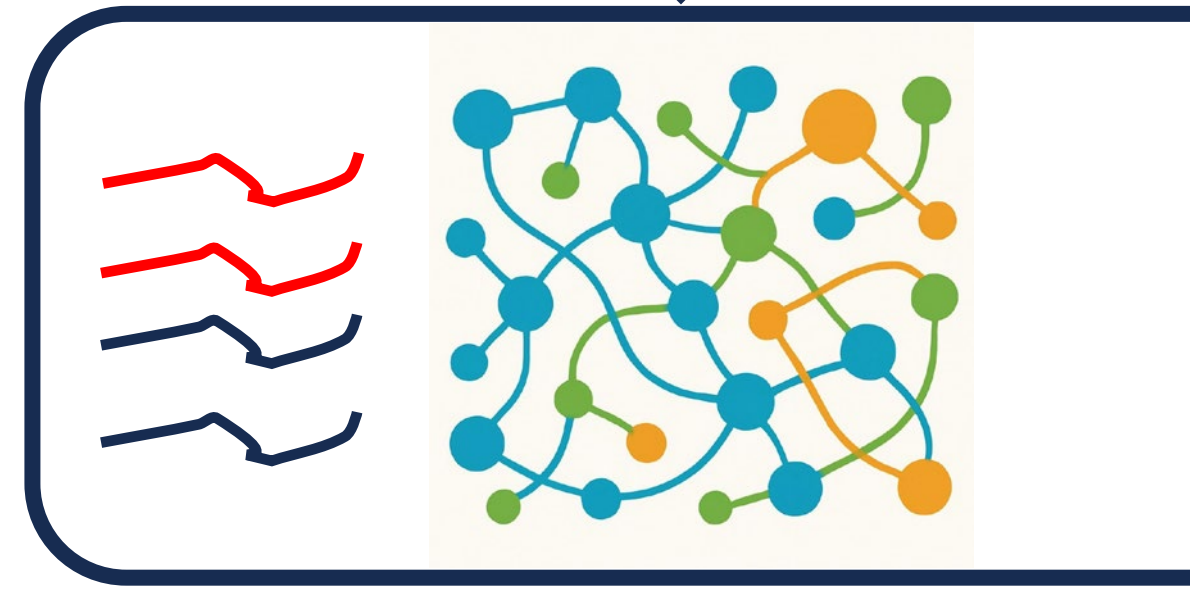

**Reproducer-replicator unit**

**Emergence of multilevel selection/ Origin of life?**

Random division

Selection at both GE and protocell levels

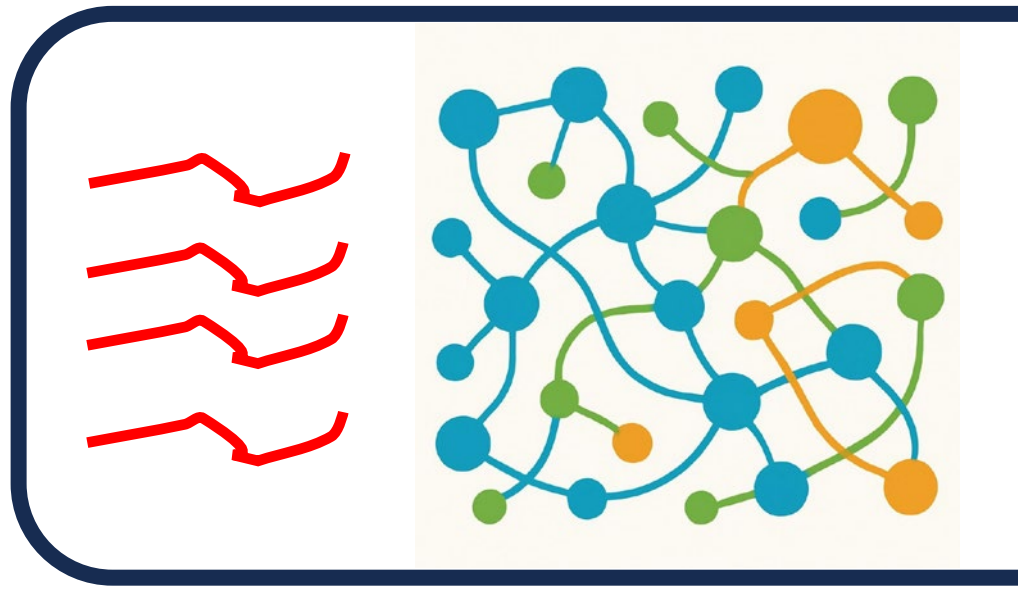

**Reproducer-replicator unit** with autonomous mutualist GE only

Selection for reduced redundancy of replication machinery. Emergence of chromosomes

Chromosome-containing prokaryote-like (proto)cell

Fig. 5